\documentclass[letterpaper, 10 pt, conference]{ieeeconf}  % Comment this line out if you need a4paper

\IEEEoverridecommandlockouts                              % This command is only needed if 
\usepackage{epsfig} % for postscript graphics files

\usepackage{amsmath} % assumes amsmath package installed
\usepackage{amssymb}  % assumes amsmath package installed
\usepackage{amsthm}
\usepackage{booktabs}
\usepackage{wrapfig}
\usepackage{mathrsfs}
\usepackage{caption}
\usepackage{subcaption}

\newtheorem{assumption}{Assumption}
\newtheorem{thm}{Theorem}

\newtheorem{remark}{Remark}
\theoremstyle{definition}

\theoremstyle{remark}

\usepackage{xcolor}
\usepackage{mathtools}

\usepackage{algorithm}
\usepackage{algorithmic}
\usepackage[algo2e]{algorithm2e}
\SetKwComment{Comment}{$\triangleright$\ }{}
\usepackage{url}

\title{\LARGE \bf
Disturbance Rejection-Guarded Learning for Vibration Suppression of Two-Inertia Systems
}

\author{Fan Zhang$^{1}$, Jinfeng Chen$^{1}$, Yu Hu$^{1}$, Zhiqiang Gao$^{1}$, Ge Lv$^{2}$, Qin Lin$^{1}$ % <-this % stops a space
\thanks{$^{1}$The authors are with the Center for Advanced Control Technologies (CACT), Cleveland State University, 2121 Euclid Avenue, Cleveland, OH 44115, USA. Corresponding author: Qin Lin, {\tt\small q.lin80@csuohio.edu}}%
\thanks{$^{2}$Ge Lv is with the Department of Mechanical Engineering, Clemson University, 105 Sikes Hall, Clemson, SC 29634, USA.}
}

\begin{document}

\maketitle
\thispagestyle{empty}
\pagestyle{empty}

%%%%%%%%%%%%%%%%%%%%%%%%%%%%%%%%%%%%%%%%%%%%%%%%%%%%%%%%%%%%%%%%%%%%%%%%%%%%%%%%
\begin{abstract}
Model uncertainty presents significant challenges in vibration suppression of multi-inertia systems, as these systems often rely on inaccurate nominal mathematical models due to system identification errors or unmodeled dynamics. An observer, such as an extended state observer (ESO), can estimate the discrepancy between the inaccurate nominal model and the true model, thus improving control performance via disturbance rejection. The conventional observer design is memoryless in the sense that once its estimated disturbance is obtained and sent to the controller, the datum is discarded. In this research, we propose a seamless integration of ESO and machine learning. On one hand, the machine learning model attempts to model the disturbance. With the assistance of prior information about the disturbance, the observer is expected to achieve faster convergence in disturbance estimation. On the other hand, machine learning benefits from an additional assurance layer provided by the ESO, as any imperfections in the machine learning model can be compensated for by the ESO. We validated the effectiveness of this novel learning-for-control paradigm through simulation and physical tests on two-inertial motion control systems used for vibration studies.

%however, the past data are retained for training to capture uncertain internal dynamics as well as patterns of external disturbance. The benefits of integrating machine learning with an observer are twofold: firstly, the learning-based estimation gains additional assurance through the observer's compensation mechanism, particularly in scenarios where the learning has not converged. Secondly, model learning facilitates a model-based ESO that demonstrates faster convergence in disturbance estimation compared to the traditional model-free ESO.  %We have validated the efficacy of this new paradigm of learning for control through physical tests of a motion control system that is used for torsional vibration study and simulations of a robotic arm. %Video: \url{https://youtu.be/h-4bwztfdlY}

\end{abstract}

\begin{keywords}
 Machine Learning, Disturbance Rejection, Extended State Observer, Model Uncertainty\\
\end{keywords}

%%%%%%%%%%%%%%%%%%%%%%%%%%%%%%%%%%%%%%%%%%%%%%%%%%%%%%%%%%%%%%%%%%%%%%%%%%%%%%%%
\section{Introduction}
%Disturbance rejection in control systems is crucial as it ensures that the system can maintain its desired performance and stability despite disturbances, making it essential for real-world applications where high precision and robust performance are required.
Vibration suppression of multi-inertia systems is critical in many engineering applications, including automotive suspensions, series elastic actuators (SEA), and various other motion control systems \cite{297904}. These systems often involve multiple inertia components with a two-inertia subsystem serving as a fundamental block, connected by flexible couplings, which leads to inherent resonance issues. This resonance can cause dynamic stresses, energy wastes, and performance degradation, therefore posing significant challenges to the systems' efficiency and stability \cite{zhao2013active,wang2024integrated}. Given the fundamental challenge of system identification and the necessity for real-time performance, it is common practice to employ a simplified or inaccurate nominal dynamic model. Consequently, the disturbances become inevitable, necessitating their rejection to achieve robust control. The disturbance includes \emph{internal} (i.e., unknown or unmodelled parts of the plant dynamics) and \emph{external} (i.e., perturbations from the outside affecting the dynamics) \cite{chen2021practical,zheng2018active}. 

The observer-based method has emerged as a promising approach to estimating the disturbance for the subsequent design of a disturbance rejection controller. Among the array of existing disturbance observers, the extended state observer (ESO) \cite{Han2009} is gaining popularity due to its simplicity in implementation. For the formulation of an ESO, the system is modeled as a simple chained integrator with a total disturbance term (also called \emph{lumped disturbance, $f$}) that includes both internal and external disturbances. The total disturbance is treated as an extended state to be estimated together with other states. The estimated disturbance can be mitigated through various means, including a simple state feedback controller or more advanced control strategies such as sliding mode control \cite{cui2017extended} and model predictive control \cite{zhang2021extended}.

It is worth noting that the traditional ESO operates in a memoryless fashion, i.e., once it estimates a disturbance and transmits it to the controller, the datum used for estimation is then discarded. However, as a control system operates, we can improve our understanding of the disturbance through collected operational data. Prior works \cite{zhang2016active,fu2016tuning} show that a model-based ESO (MB-ESO), which utilizes prior model information about the disturbance (such as a detailed dynamic model obtained through system identification), tends to exhibit reduced sensitivity to noise when compared to a model-free ESO (MF-ESO) that assumes a simple chained integrator as a nominal model. In order to circumvent the need for extensive system identification and maximize the utilization of disturbance information, we propose to leverage machine learning (ML), which has powerful capacities for nonlinear optimization, to memorize and generalize the past estimations from the ESO as a \emph{feedforward} estimation of the disturbance. The learning component is expected to capture the internal dynamics as well as patterns of external disturbances.

\cite{8689051,9736588,zhang2022improving} combine ESO with iterative learning control (ILC) for repetitive control tasks. Our approach focuses on general control tasks rather than just the repetitive ones. In addition, we assume that system dynamics, as well as disturbances, are unknown and not necessarily repetitive. In \cite{kicki2022tuning}, a neural network is utilized to tune the parameters of ESO rather than explicitly learning the disturbance. Other learning-for-control approaches such as \cite{NeuralLander} employ neural networks to capture discrepancies between a nominal model $\hat{F}(x_k, u_k)$ and the true model $F(x_k, u_k)$. Since the state of the true model is unknown, the measured next state $x_{k+1}$ is used to update the error model represented by the neural network. However, these methods always assume full-state information is available. In addition, when the learning performance falls short of expectations, it may result in suboptimal performance for subsequent model-based controllers. In contrast, our approach represents a novel paradigm that aims at learning the total disturbance with the help of output measurements instead of true values for states. Furthermore, our paradigm includes a correction mechanism for cases where the learning component fails to accurately capture the disturbances. The residual total disturbance, i.e., the remainder excluding the disturbance already estimated by the learning component, will be estimated by a conventional ESO in a \emph{feedback} correction manner. Through this seamless integration, even when the learning-based estimation struggles to converge effectively, we can leverage the ESO for feedback correction, thereby adding an extra layer of robustness and assurance to the system.

%\textcolor{blue}{Using bold font is not my common practice, but if this is something that you always do then please ignore my comment. }

\begin{figure}[htbp]
    \centering
    \includegraphics[width=0.8\linewidth]{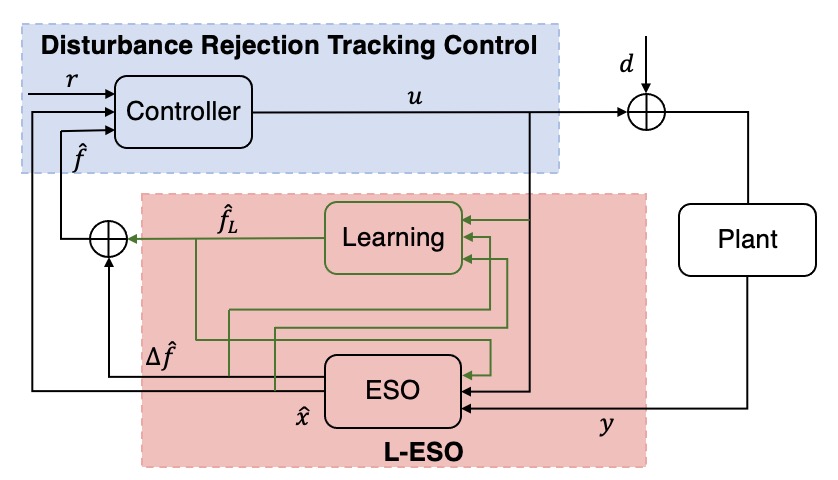}
    \caption{The proposed framework in this paper, where the red and the blue blocks represent the L-ESO and the disturbance rejection tracking controller, respectively. Once the total disturbance is estimated, the tracking controller will be able to reject disturbance.}
    \label{fig:LESO}
\end{figure}

In our new framework, as visualized in Fig. \ref{fig:LESO}, we refer to the learning-enabled extended state observer as L-ESO. The estimation $\hat{f}$ of the true total disturbance $f$ consists of $\hat{f}_{L}$ and $\Delta \hat{f}$, which are from the learning component and the ESO, respectively. First, ESO uses the information of control $u$ and observation $y$ to estimate the system's states $\hat{x}$ and the residual disturbance $\Delta \hat{f}$. Second, ESO's estimation, including $\hat{x}$ and $u$ are fed as input to the learning component for learning a regression model. The learning component carries out the feedforward estimation $\hat{f}_{L}$, after which an online optimization iteratively minimizes the difference between $\hat{f}_{L}$ and $\hat{f}$, allowing the learning component to approximate the total disturbance accurately. In situations where imperfect learning introduces errors, the ESO serves as an additional layer to rectify.

The contributions of our work are summarized as follows:
\begin{itemize}
  \item We propose a novel framework that combines ML and ESO for feedforward estimation and feedback correction for a general disturbance rejection tracking control task. Compared with existing learning-for-control frameworks, we estimate states and disturbances in a unique way. We also have an extra error correction mechanism for the learning component.
   \item The learning component serves as an add-on to existing ESO-based control architecture. As shown in Fig. \ref{fig:LESO}, only a learning component and a few connections (in green) are introduced. The advantage of our modular design is two-fold: 1) no need to change the existing framework; 2) users can customize the learning components by choosing any appropriate machine learning model.
   %\item \textcolor{red}{Conventional ESO suffers from bandwidth limits causing the gap between $f$  and $\Hat{f}$. By gradually adding model information, such gap can be narrowed eventually, leading to better control performance.}
  %\item We demonstrate that in simulations and physical experiments, at least, L-ESO is comparable to MB-ESO; when the external disturbance has some regular patterns, L-ESO is expected to even outperform MB-ESO. %the proposed L-ESO is comparable or even beyond the model-based ESO that is assisted by a fully known model. %Additionally, learning makes it more robust against changes to plant dynamics or external disturbances.
  \item Our learning and estimation are real-time and online. We showcase the efficacy of our framework through simulations and a real-world two-inertia testbed as a fundamental block for a multi-inertia system. %\Additionally, we conduct simulations of attitude and altitude control for a quadrotor unmanned aerial vehicle (UAV) that is a Multi-Input Multi-Output (MIMO) system with coupled dynamics and disturbances.
\end{itemize}

The remainder of this paper is structured as follows. We first go through the preliminaries in Sec. \ref{sec:background}. Then, we construct our framework in Sec. \ref{sec:Learning based ESO}. Simulation results of the two-mass-spring benchmark system are presented in Sec. \ref{sec:simulation}, followed by the hardware experiments of a torsional plant in Sec. \ref{sec:hardware}. Finally, we conclude our work and discuss possible future research directions in Sec. \ref{sec:conclusion}.

%\section{Related Works}
\label{sec:related}

\section{Preliminary}
\label{sec:background}
% \subsection{System dynamics and Extended State Observer}

The multi-inertia system can be represented as the sum of a nominal part and a nonlinear time-varying part:
\begin{equation}\label{eq_1}
    \begin{cases}
        \dot{\Bar{x}}(t) = A_0 \Bar{x}(t)+B_0 u(t) + E_0 f(x(t), d(t), t)\\
        y=C_0 \Bar{x}
    \end{cases}
\end{equation}
where $\Bar{x} \in \mathbb{R}^n$ is the state vector, $u\in \mathbb{R}$ is a control input, $y\in\mathbb{R}$ is a measured output, and $f: \mathbb{R}^{n+1} \times [0, \infty] \rightarrow \mathbb{R}$ is an unknown function representing the time-varying uncertainty, which contains external disturbance $d(t) \in \mathbb{R}$, unmodeled dynamics, and parameter uncertainty. Terms $A_0$, $B_0$, $E_0$ and $C_0$ are real and known matrices with appropriate dimensions. For the particular case of a two-inertial system with $n=4$, meaning two states for each inertial position/angular and velocity/angular velocity, please refer to the details in the example in Sec. \ref{sec:simulation}. The justification of classifying \eqref{eq_1} as a nonlinear time-varying system can be found in \cite{guo2011convergence,bai2019observers}.

Traditionally, an ESO is established for a system in a chained integrator form \cite{Han2009}. However, in our most recent work \cite{chen2022relationship}, we have significantly expanded the applicability scope of ESO and rigorously proved that for a general system \eqref{eq_1}, given that Assumption \ref{asm1} and the Assumption \ref{asm3} are satisfied, an ESO can be established to estimate $f$ by releasing the chained integrator form requirement.

\begin{assumption}\label{asm1}
  $(A_0, C_0)$ is observable.
\end{assumption}
% \begin{assumption}\label{asm2} 
%  $(A_0,B_0, C_0)$ has no invariant zeros.
% \end{assumption}
\begin{assumption}\label{asm3}
   $(A_0,E_0, C_0)$ has no invariant zeros.
\end{assumption}

For system \eqref{eq_1}, under the Assumptions \ref{asm1}, and \ref{asm3} , there exists a matrix 
\begin{equation}\label{}
    S=\begin{bmatrix}
        C_0 \\
        C_0 A_0\\
        \vdots\\
        C_0 A_0^{n-1}
    \end{bmatrix}
\end{equation}
such that 

\begin{equation}\label{eq:transform}
 \begin{split}
    \Bar{A}_0=SA_0S^{-1}&=\begin{bmatrix}
        0&1&\dots&0 \\
        \vdots&\ddots&\ddots&\vdots \\
        0&\dots&0&1 \\
        -a_0&-a_1&\dots&-a_{n-1} \\
    \end{bmatrix} \\
      \Bar{B}_0=SB_0&=\begin{bmatrix}
    0 &0&\dots&b
    \end{bmatrix}^T\\
     \Bar{C}_0=C_0S^{-1}&=\begin{bmatrix}
        1&0&\dots&0 \\
    \end{bmatrix}\\
  \Bar{E}_0=SE_0&=\begin{bmatrix}
        0 &0&\dots&1
    \end{bmatrix}^T\\
\end{split}
\end{equation}
form the following new system

\begin{equation}\label{system2}
    \begin{cases}
        \dot{x}=\bar{A}_0 x+\bar{B}_0 u + \bar{E}_0 f\\
        y=\bar{C}_0 x
    \end{cases}
\end{equation}

The readers are referred to \cite{chen2022relationship} for more details on the matrix transformation. The new system \eqref{system2} has an observable canonical form such that an ESO can be established for estimating $f$. 
% In this paper, only matched disturbance is considered in a SISO system for simplicity, i.e., \(f\) and \(u\) are in the same channel, which is implied in Assumption \ref{asm2}.

\begin{remark}
Assumption \ref{asm3} is equivalent to the following conditions. The proof can be found in \cite{chen2022relationship}.\\

   % \centerline{$C_0B_0=0,C_0A_0B_0=0,\dots, C_0A_0^{n-2}B_0=0,C_0A_0^{n-1}B_0 \neq 0$ }
   \centerline{$C_0E_0=0,C_0A_0E_0=0,\dots, C_0A_0^{n-1}E_0 \neq 0$ }
\end{remark}

According to whether or not the system dynamics are available, we have the following two variants of ESO:
\subsection{MB-ESO} If the model information, i.e., $-a_0, -a_1, \cdots,-a_{n-1},b$, in  matrix $\bar{A}_0$ and $\bar{B}_0$ is available, we have
\begin{equation}\label{eq:mbplant}
\begin{aligned}
\dot{x}=&\underbrace{\begin{bmatrix}
        0&1&\dots&0 \\
        \vdots&\ddots&\ddots&\vdots \\
        0&\dots&0&1 \\
        -a_0&\dots&\dots&-a_{n-1} \\
    \end{bmatrix} }_{\Bar{A}_{0,MB}}
        x
    +\underbrace{\begin{bmatrix}
        0 \\
        0\\
        \vdots\\
        b
    \end{bmatrix}}_{\Bar{B}_{0,MB}} u+
    \underbrace{\begin{bmatrix}
        0 \\
        0\\
        \vdots\\
        1
    \end{bmatrix} }_{\Bar{E}_0}
    \underbrace{d}_{f}
     \end{aligned}
\end{equation}

The total disturbance can be represented as:
\begin{equation}\label{eq:total-d-MB}
        f=d\\
\end{equation}
where $d$ is the external disturbance, $b$ is the true control gain.

\subsection{MF-ESO} If the model information, i.e., $-a_0, -a_1, \cdots,-a_{n-1},b$, in  matrix $\bar{A}_0$ and $\bar{B}_0$, is not available, we have
\begin{equation}\label{eq:mfplant}
\begin{array}{r@{}l}
\dot{x}=&\underbrace{\begin{bmatrix}
        0&1&\dots&0 \\
        \vdots&\ddots&\ddots&\vdots \\
        0&\dots&0&1 \\
        0&0&\dots&0 \\
    \end{bmatrix} }_{\Bar{A}_{0,MF}}
        x
    +\underbrace{\begin{bmatrix}
        0 \\
        0\\
        \vdots\\
        b_0
    \end{bmatrix}}_{\Bar{B}_{0,MF}} u+\\
     &   
     \underbrace{\begin{bmatrix}
        0 \\
        \vdots\\
        1
    \end{bmatrix}}_{\Bar{E}_0}
    \underbrace{  (-a_0x_1-\dots-a_{n-1}x_n+(b-b_0)u+d)}_{f}
    
    \end{array}
\end{equation}
where $-a_0x_1-\dots-a_{n-1}x_n+(b-b_0)u$ is the internal disturbance (unknown/unmodelled dynamics), $b_0$ is the nominal control gain, and $d$ is the external disturbance. In such a case, the total disturbance becomes:

\begin{equation}\label{eq:total-d-MF}
   f=-a_0x_1-\dots-a_{n-1}x_n+ (b-b_0) u+d\\
\end{equation}

ESO treats the total disturbance $f$ as an extended state, such that a Luenberger observer can be designed to estimate both the original system state $x$ and the total disturbance $f$. The augmented dynamic system is as follows:
\begin{equation}\label{eq_2}
    \begin{cases}
            \begin{bmatrix}
        \dot{x} \\
        \dot{f}
    \end{bmatrix}=A \begin{bmatrix}
        x\\
        f
    \end{bmatrix}+B u + E \dot{f}\\
        y=C x
    \end{cases}
\end{equation}
 where \(A=\begin{bmatrix}
    \Bar{A_0} & \Bar{E_0}\\
    0_{1\times n} & 0
\end{bmatrix}_{(n+1)\times(n+1)}\), 
\(B=\begin{bmatrix}
    \Bar{B_0}\\
    0
\end{bmatrix}_{(n+1)\times 1}\), 
\(C=[\Bar{C_0}, 0]_{1\times (n+1)}\), 
\(E=[0, \cdots, 0, 1]_{(n+1)\times 1}^T \). 

The Luenberger observer has the following form:

\begin{equation}\label{eq_3}
    \begin{bmatrix}
        \dot{\hat{x}} \\
        \dot{\hat{f}}
    \end{bmatrix} = A \begin{bmatrix}
        \hat{x}\\
        \hat{f}
    \end{bmatrix} + B u+L\left( y-C \begin{bmatrix}
        \hat{x}\\
        \hat{f}
    \end{bmatrix}\right)
\end{equation}
where \(\hat{x}\) and \(\hat{f}\) are estimations of \(x\) and \(f\), \(L\) is the observer gain. We have the following estimation error dynamics:
% The error dynamics of observer \eqref{eq_3} are:
\begin{equation}\label{}
        \dot{e}=(A-LC) e+E \dot{f}\\
\end{equation}
where $
    e= \begin{bmatrix}
        x-\hat{x} & f-\hat{f}\\
    \end{bmatrix}^T 
$.

\begin{thm}\label{thm1}
Under Assumption \ref{asm1} and Assumption \ref{asm3}, the eigenvalues $A-LC$ can be placed at the left side of the plane to make the estimation converge \cite{chen2022relationship,bai2019observers}.
\end{thm}
All eigenvalues can be placed at $-\omega_o$, which is called the observer bandwidth of ESO \cite{gao2003scaling}.

\section{Learning-Enabled ESO}\label{sec:Learning based ESO}
%In this section, we will elaborate on our proposed L-ESO.

%As shown in Eq. \eqref{eq_3}, ESO takes the system's observation $y$ and control $u$ as input and carries out the estimation of states $\hat{x}$ and the total disturbance $\hat{f}$.
%The choice of the matrix $\bar{A}_0$ and $\bar{B}_0$ distinguishes 

The model-based ESO in \eqref{eq:mbplant} and the model-free ESO in \eqref{eq:mfplant} can be further expanded as follows:
\begin{equation}\label{eq:LearningA}
\begin{array}{r@{}l}
&\underbrace{\begin{bmatrix}
  \underbrace{\begin{matrix}
   0&1&\dots&0 \\
        \vdots&\ddots&\ddots&\vdots \\
        0&\dots&0&1 \\
        -a_0&\dots&\dots&-a_{n-1} \\
  \end{matrix}}_{\Bar{A}_{0,MB}}
  & \Bar{E_0} \\
   0_{1\times n} & 0
\end{bmatrix} }_{A_{MB}}
\begin{bmatrix}
        \hat{x}\\
        \hat{f}
    \end{bmatrix}
    +
    \underbrace{\begin{bmatrix}
        0 \\
        \vdots\\
       \underbrace{ b_0+ b-b_0}_{\Bar{B}_{0,MB}}
        \\0
    \end{bmatrix} }_{B_{MB}}u
    =\\
    
  &  \underbrace{\begin{bmatrix}
  \underbrace{\begin{matrix}
   0&1&\dots&0 \\
        \vdots&\ddots&\ddots&\vdots \\
        0&\dots&0&1 \\
        0&\dots&\dots&0 \\
  \end{matrix}}_{\Bar{A}_{0,MF}}
  & \Bar{E_0} \\
   0_{1\times n} & 0
\end{bmatrix} }_{A_{MF}}
\begin{bmatrix}
        \hat{x}\\
        \hat{f}
    \end{bmatrix}
        +
        \underbrace{\begin{bmatrix}
        0 \\
        \vdots\\
       \underbrace{ b_0}_{\Bar{B}_{0,MF}}
        \\0
    \end{bmatrix} }_{B_{MF}}u

        +\\
      &   \begin{bmatrix}
        \Bar{E}_0\\
        0
    \end{bmatrix} (-a_0x_1\dots-a_{n-1}x_n+ (b-b_0)u)
    \end{array}
\end{equation}

\begin{remark}
By incorporating model information, MF-ESO becomes equivalent to MB-ESO.     
\end{remark}

\begin{remark}
The motivation for proposing the learning component can be justified in that the model information is learnable to facilitate the incorporation of model information.
\end{remark}

%Thus, the ESO Eq. \ref{eq_3} can be modified as follows since the learning block shares the job with ESO.

\begin{remark}
    The learning component is even possible to learn the external disturbance together with the internal disturbance to be incorporated.
\end{remark}

Since the learning component has a feedforward estimation $\hat{f}_L$ for the total disturbance, ESO can serve as a feedback correction to estimate the residual total disturbance as $\Delta\hat{f}$. The combination of the feedforward estimation and the feedback correction is realized as follows: 
\begin{equation}\label{eq.leso}
    \begin{bmatrix}
        \dot{\hat{x}} \\
        \dot{\Delta\hat{ f}}
    \end{bmatrix} = A \begin{bmatrix}
        \hat{x}\\
        \Delta\hat{f}
    \end{bmatrix} + B u+L\left( y-C \begin{bmatrix}
        \hat{x}\\
        \Delta\hat{f}
        \end{bmatrix}\right)+
         \begin{bmatrix}
        \Bar{E}_0\\
        0
    \end{bmatrix}\hat{f}_{L}
\end{equation}

Since the learning component is expected to capture the unknown dynamics, we employ a model-free ESO, see Fig. \ref{fig:LESO}.
% \begin{figure}[htbp]
%     \centering
% \includegraphics[width=0.75\linewidth]{Images/LearningESOdetail.pdf}
%     \caption{Detailed illustration for the combination of the learning component and ESO.}
%     \label{fig:Leso}
% \end{figure}
The learning block in Fig. \ref{fig:LESO} is a function $h_\theta(x,u)$ parameterized by $\theta$. 
To learn the total disturbance (see \eqref{eq:total-d-MF}), we establish a mapping from the input ($\hat{x}$ estimated by ESO and control input $u$) to the output $\hat{f}$, where $\hat{f}=\hat{f}_{L}+\Delta\hat{f}$. The total disturbance estimation consists of two parts: 1) the feedforward estimation from the learning component $\hat{f}_{L} = h_\theta(\hat{x}, u)$; 2) feedback correction for the residual disturbance $\Delta\hat{f}$ by an MF-ESO. To optimize the parameters of the machine learning model, a general regression problem is formulated using the following cost function:
\begin{equation}\label{eq:cost}
        J(\theta)= \frac{1}{2} \sum_{i=1}^{n} (h_{\theta}(\hat{x}^i,u^i) -\hat{f}^i)^2
\end{equation}
where $n$ is the size of the training data. The details are in Alg. 1. When the batch is not yet filled, we run the MF-ESO (see Line 7-14, the learning component does not return optimized parameters). %Note that in practice, the true value of the total disturbance $f$ is never known, so it can not be directly used as a label in supervised learning. However, thanks to the convergence of ESO (see Theorem \ref{thm1}), \textcolor{red}{we can iteratively use the last step's estimation $\hat{f}$ as an available label to optimize Eq. \eqref{eq:cost}}.

\begin{algorithm}\label{alg:Leso}
\setcounter{AlgoLine}{0}
\LinesNumbered
\caption{L-ESO}
\KwIn{ Control input $u$, system output $y$, learning rate $\alpha$, batch size $n$, maximum running time $N_{max}$ }
\KwOut{Total disturbance $\hat{f}$}
Initialize:

machine learning input batch $\mathcal{I}^0=\emptyset$

disturbance estimation by ESO batch $\Delta \mathcal{F}^0=\emptyset$ 

machine learning output batch $\mathcal{F_L}^0=\emptyset$ 

machine learning model parameter $\theta$

machine learning output $\hat{f}_L^0=0$ 

\For{ $i=1$ to $n$}{
Get $\hat{x}^i$ and $\Delta \hat{f}^i$ by running L-ESO \hfill \Comment{see \eqref{eq.leso}}

Compute $u^i$  \hfill \Comment{see \eqref{eq=control}}

$\mathcal{I}^i := [ \mathcal{I}^{i-1} , [\hat{x}_1^i,\hat{x}_2^i,\dots,\hat{x}_n^i,u^i,1]^T]$

$\Delta \mathcal{F}^i := [\Delta \mathcal{F}^{i-1} , \Delta \hat{f}^i]$

$\mathcal{F_L}^i := [\mathcal{F_L}^{i-1} , 0]$  \hfill \Comment{append data into three batches}

$\hat{f}_L^{i}=0$
}

\For{$i=n$ to $N_{max}$}{

Get $\hat{x}^i$ and $\Delta \hat{f}^i$ by running L-ESO   \hfill \Comment{see \eqref{eq.leso}}

Update $\mathcal{I}^i$ \hfill \Comment{pop oldest datum, push new datum}

Update $\Delta \mathcal{F}^i$ \hfill \Comment{pop oldest datum, push new datum}

Update $\theta^{i}$ \hfill \Comment{According to \eqref{eq:cost}}

$\mathcal{F_L}^i=h_{\theta^{i}}(\mathcal{I}^{i})$

$\hat{f}_{L}^{i}=h_{\theta^{i}}(x^i)$

$\hat{f}^{i}=\hat{f}_{L}^{i}+\Delta \hat{f}^i$  \hfill \Comment{compute total disturbance}

Compute $u^i$  \hfill \Comment{see \eqref{eq=control}}
}

\end{algorithm}

Our framework has superior modularity. The design of the ESO is just a conventional model-free convention. We only need to use the estimation from ESO to drive the training of our learning component. First, the learning component can serve as an add-on to existing ESO-based control architecture by just adding a few connections. Second, the learning component is so flexible that users can customize it by choosing appropriate machine learning models, e.g., linear, non-linear, parametric, non-parametric, etc. %Since the learning block is independent, this approach can be implemented both offline or online, with offline implementation being safer and controllable; the online approach is more robust against changes to plant dynamics or external disturbances. Moreover, by utilizing various ML models, it can be applied to both linear and nonlinear systems.

\section{Simulation Results}
\label{sec:simulation}
%In this section, we conduct simulation and physical experiments for a comprehensive performance evaluation for L-ESO. We also compare L-ESO with two baseline approaches, i.e., MB-ESO and MF-ESO \cite{zhang2016active}. 

\subsection{Two-Mass-Spring Problem Formulation}
\begin{figure}[htbp]
    \centering
    \includegraphics[width=0.8\linewidth]{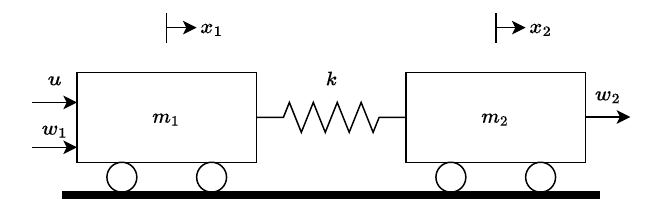}
    \caption{Two-mass-spring system with uncertain parameters}
    \label{fig:twomass}
\end{figure}

Fig. \ref{fig:twomass} depicts a schematic of a two-mass-spring system, which is from a well-known benchmark control problem \cite{wie1992benchmark}. The system includes two masses: $m_1$ and $m_2$, which can slide freely over a horizontal surface without friction. Note that it has been proved that a non-friction setting is more challenging for a controller design \cite{zhang2016active}. The masses are connected by a light horizontal spring with a spring constant $k$. The system is subject to two external disturbance forces $w_1$ and $w_2$, which act on masses $m_1$ and $m_2$, respectively. The control signal $u$ is the force applied to mass $m_1$. Both the positions of mass $m_1$ and mass $m_2$ are measured, and either one can be used as an output to be controlled.

The states of the two-mass-spring system are defined as the displacements and velocities of the two masses. Specifically, the displacement and velocity of mass $m_1$ are $x_1$ and $x_3$, respectively, while the displacement and velocity of mass $m_2$ are $x_2$ and $x_4$, respectively. The dynamics of the system can be represented in the following state-space form:
\begin{equation}\label{eq:two-mass}
    \begin{split}
    \begin{bmatrix}
        \dot{x}_1\\
        \dot{x}_2\\
        \dot{x}_3\\
        \dot{x}_4
    \end{bmatrix} &=
    \begin{bmatrix}
        0& 0 & 1&0\\
         0& 0 & 0&1\\
          -\frac{k}{m_1}&  \frac{k}{m_1} & 0 &0\\
            \frac{k}{m_2}&  -\frac{k}{m_2} & 0 &0\\
    \end{bmatrix}
    \begin{bmatrix}
        x_1\\
        x_2\\
        x_3\\
        x_4\\
    \end{bmatrix}\\
    &+\begin{bmatrix}
        0\\
        0\\
        \frac{1}{m_1}\\
        0\\
    \end{bmatrix}(u+w_1)
    +\begin{bmatrix}
        0\\
        0\\
        0\\
         \frac{1}{m_2}
    \end{bmatrix}w_2\\
    y&=\begin{bmatrix}
        c_1 & c_2 & 0 & 0
    \end{bmatrix} \begin{bmatrix}
        x_1 & x_2 & x_3 &x_4\\
    \end{bmatrix}^T
     \end{split}
\end{equation}

%Imagine a common damper system as our experimental setting,

A time-varying unknown external disturbance $w_2$ is from the mass $m_2$, control needs to be conducted on $m_1$ to allow $x_2$ track any desired trajectory. For the output $y$, i.e., $x_2$, a chained integrator system is derived by taking the derivatives of the output four times. The input and disturbance are in the last channel of this fourth-order system with $b = \frac{k}{m_1m_2}$:  %\textcolor{red}{Assuming the disturbance is all coming from $w_2$, the transfer function of the system can be written as:}
\begin{equation}\label{eq:4order}
y^{(4)}=-k\frac{m_1+m_2}{m_1m_2}\ddot{y}+\frac{k}{m_1m_2}w_2+\frac{1}{m_2}\ddot{w}_2+bu
\end{equation}
%where .
\subsection{ESO design}
\label{sec:esodesign}
The states in the system are:
\begin{equation}\label{}
x=\begin{bmatrix}
    y&
    \dot{y}&
    \ddot{y}&
    \dddot{y}
\end{bmatrix}^T
\end{equation}
The state-space description of the system is 
\begin{equation}\label{}
    \begin{cases}
            \begin{bmatrix}
        \dot{x} \\
        \dot{f}
    \end{bmatrix}=A \begin{bmatrix}
        x\\
        f
    \end{bmatrix}+B u + E \dot{f}\\
        y=C x
    \end{cases}
\end{equation}

\subsubsection{Model-free ESO}
The state-space model is:

 \(A_{MF}=\begin{bmatrix}
        0& 1 & 0&0&0\\
         0& 0 & 1&0&0\\
          0& 0 & 0&1&0\\
           0& 0 & 0&0&1\\
           0& 0 & 0&0&0\\
\end{bmatrix}\), 
\(B=\begin{bmatrix}
    0\\
    0\\
    0\\
    b_0\\
    0
\end{bmatrix}\), 
\(C=\begin{bmatrix}1 &0& 0& 0& 0\end{bmatrix}\), 
\(E=\begin{bmatrix}0 &0&0&0&1\end{bmatrix}^T\).
As we can see, the model-free design assumes unknown dynamics, such that the total disturbance $f$ can be represented as:
\begin{equation}\label{eq:disturbance-twomass-MFESO}
        f=-k\frac{m_1+m_2}{m_1m_2}\ddot{y}+\frac{k}{m_1m_2}w_2+\frac{1}{m_2}\ddot{w}_2+ (b - b_0 ) u
\end{equation}
where $-k\frac{m_1+m_2}{m_1m_2}$ is the model parameter information, $b_0$ is the nominal control gain. We have
\begin{equation}\label{eq:two-mass-y4}
y^{(4)} = f + b_0 u
\end{equation}
 where everything besides $b_0 u$ is considered as total disturbance (see \eqref{eq:4order}). It can be validated that such a system satisfies Assumptions 1, 2, and 3. Therefore, an ESO can be designed for the estimation of $f$, see \eqref{eq_3}. %Note that to get such a chained integrator representation, you can also start with the system \eqref{eq:two-mass} and then use the transformation in Eq. \eqref{eq:transform} to get the same form.

The observer gain is chosen where all the eigenvalues of $A_{MF}-LC$ are placed at $-\omega_o$ \cite{gao2003scaling}, i.e., $L_{MF}=[5\omega_o \quad 10\omega_o^2 \quad 10\omega_o^3 \quad 5\omega_o^4 \quad \omega_o^5]$.

\subsubsection{Model-based ESO}
The model-based design has the following state-space representation:

 \(A_{MB}=\begin{bmatrix}
        0& 1 & 0&0&0\\
         0& 0 & 1&0&0\\
          0& 0 & 0&1&0\\
           0& 0 & -k\frac{m_1+m_2}{m_1m_2}&0&1\\
           0& 0 & 0&0&0\\
\end{bmatrix}\), 
\(B=\begin{bmatrix}
    0\\
    0\\
    0\\
    b_0\\
    0
\end{bmatrix}\), 
\(C=\begin{bmatrix}1 &0& 0& 0& 0\end{bmatrix}\), 
\(E=\begin{bmatrix}0 &0&0&0&1\end{bmatrix}^T\). In contrast to the above-mentioned model-free design, such a system tries to leverage the prior knowledge of the dynamic model, by assuming $-k \frac{m_1+m_2}{m_1 m_2}$ is known (see \eqref{eq:4order}). In this case, the total disturbance becomes:

\begin{equation}\label{eq:disturbance-twomass-MBESO}
        f=\frac{k}{m_1m_2}w_2+\frac{1}{m_2}\ddot{w}_2+
         (b - b_0 ) u
\end{equation}
such that $y^{(4)}=-k\frac{m_1+m_2}{m_1m_2}\ddot{y}+f+b_0 u$

The observer gain is chosen where all eigenvalues of $A_{MB}-LC$ are placed at $-\omega_o$ \cite{gao2003scaling}. Let $a=-k\frac{m_1+m_2}{m_1m_2}$, the coefficients of $L_{MB}$ are listed in Table \ref{table}.

\begin{table}[htbp]
\begin{center}
\begin{tabular}{ c c } 
 \toprule
 Parameters & Values \\
 \midrule
 $L_{MB, 1}$ & $5\omega_o$ \\ 
 $L_{MB, 2}$ & $a+10\omega_o^2$ \\ 
 $L_{MB, 3}$ & $5a\omega_o+10\omega_o^3$ \\ 
 $L_{MB, 4}$ & $a^2+10a\omega_o^2+5\omega_o^4$  \\ 
 $L_{MB, 5}$ & $5a^2\omega_o+10a\omega_o^3+\omega_o^5$ \\ 
\bottomrule 
\end{tabular}
\caption{\label{table}coefficients of $L_{MB}$}
\end{center}
\end{table}
% \begin{align*} 
% L_{model\_based}=[5\omega_o \quad a+10\omega_o^2 \quad 5a\omega_o+10\omega_o^3 \quad \\a^2+10a\omega_o^2+5\omega_o^4 \quad 5a^2\omega_o+10a\omega_o^3+\omega_o^5]
% \end{align*}
\subsubsection{L-ESO}
As shown in \eqref{eq:disturbance-twomass-MFESO}, the internal disturbance has a linearly structured mapping between the input (state and control)  and the output (disturbance). Therefore, a linear regression model is a reasonable choice for the learning component, with $h_\theta(\cdot)=\theta^T \begin{bmatrix}
    \hat{x}_1&
    \hat{x}_2&
    \hat{x}_3&
    \hat{x}_4&
    u&
    1
\end{bmatrix}^T$. Note that as we mentioned before, the learning model is flexible to be linear, nonlinear, parametric, non-parametric, etc. Our contribution is not about the complexity of the learning model but the novel design to seamlessly combine machine learning models with an ESO. A batch gradient descent method is used for optimizing the cost function. In our experiments, we initialize $\theta$ with all zeros. %except a reasonable nominal control gain $b_0$. It has been reported in the literature that the error of the nominal control gain can not go beyond two times the true value for the capacity limit for ADRC. Note that our learning phase is also closed-loop, therefore the ADRC should function under a reasonable working condition.%The $A,B,C,E$ matrix is the same as the model free version.

\subsection{Controller Design}
\label{sec:controldesign}
%Follow the conventional active disturbance rejection control (ADRC) method \cite{Han2009}, rewrite the system \eqref{eq:4order} as 
%\begin{equation}\label{yf}
%y^{(4)}=f+bu
%\end{equation}
%where $b_0=\frac{k}{m_1m_2}$ \qin{$b_0$?}is the control input gain of the system. Then

The control law for the system \eqref{eq:two-mass-y4} can be designed as:

\begin{equation}\label{eq=control}
u=\frac{-\hat{f}+u_0}{b_0}
\end{equation}
such that
\begin{equation}
y^{(4)}=u_0
\end{equation}
It can be controlled by a state feedback controller
\begin{equation}
%\begin{split}
u_0 =-K\hat{x} = k_1(r-\hat{x}_1)-k_2\hat{x}_2-k_3\hat{x}_3-k_4\hat{x}_4\\
%&=k_1(r-\hat{x}_1)-k_2\hat{x}_2-k_3\hat{x}_3-k_4\hat{x}_4-k_5\hat{x}_5
%\end{split}
\end{equation}
 with a control gain $K=\begin{bmatrix}
   \omega_c^4 &
    4\omega_c^3&
    6\omega_c^2&
    4\omega_c
\end{bmatrix}$,
where $\omega_c$ is the close-loop natural frequency \cite{gao2003scaling}.

\subsection{Simulation Results}
%The simulation results of the proposed method (L-ESO) with comparisons to the model-free (MF-ESO) and model-based (MB-ESO) approaches in MATLAB\textsuperscript \textregistered Simulink are presented. 

The system parameters are taken from the benckmark problem  \cite{wie1992benchmark}, i.e.,  $m_1 = m_2 = 1$ kg, $k = 1$ N/m, \(c_1=0\), \(c_2=1\). Tracking a desired trajectory for the position of mass $m_2$ is the control objective. A sinusoidal wave with a frequency of 1 rad/s and amplitude 1 is applied in the training phase for L-ESO.  After 110 seconds, a step reference is given to all three approaches. A band-limited white noise with noise power $10^{-12}$ is added at the system output side. A sinusoidal external disturbance with frequency $\pi/10$ rad/s is applied on $m_2$ as $w_2$ starting at 150 s. The learning algorithm is running online. The learning phase is designed to emulate the typical operational scenarios of the machine under general conditions, whereas the step response is employed to assess and compare the tracking performance. All the control parameters are set identically for fair comparison.

%The controller and observer under test are described in \ref{sec:controldesign} and \ref{sec:esodesign}, with
The controller bandwidth $\omega_c$ and the observer bandwidth $\omega_o$ are set to 1 rad/s and 10 rad/s, respectively. The control gain is set to 1. All three approaches share such same settings for fair comparison.
\begin{figure}[htbp]
    \centering
    \includegraphics[width=0.8\linewidth]{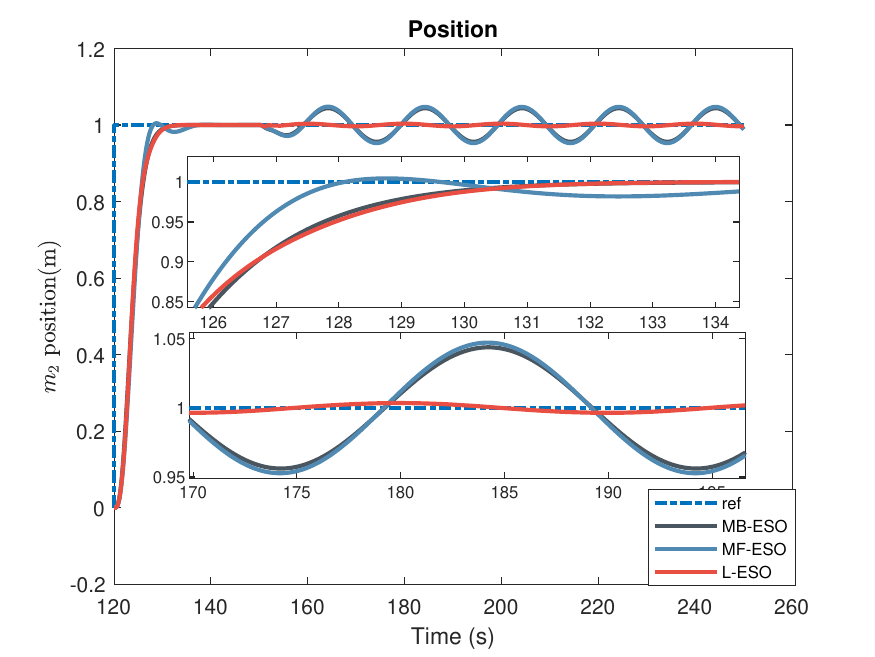}
    \caption{Tracking performance for MB-ESO, MF-ESO, L-ESO plotting from 120s.}
    \label{fig:position}
\end{figure}

\begin{figure}[htbp]
    \centering
    \includegraphics[width=0.8\linewidth]{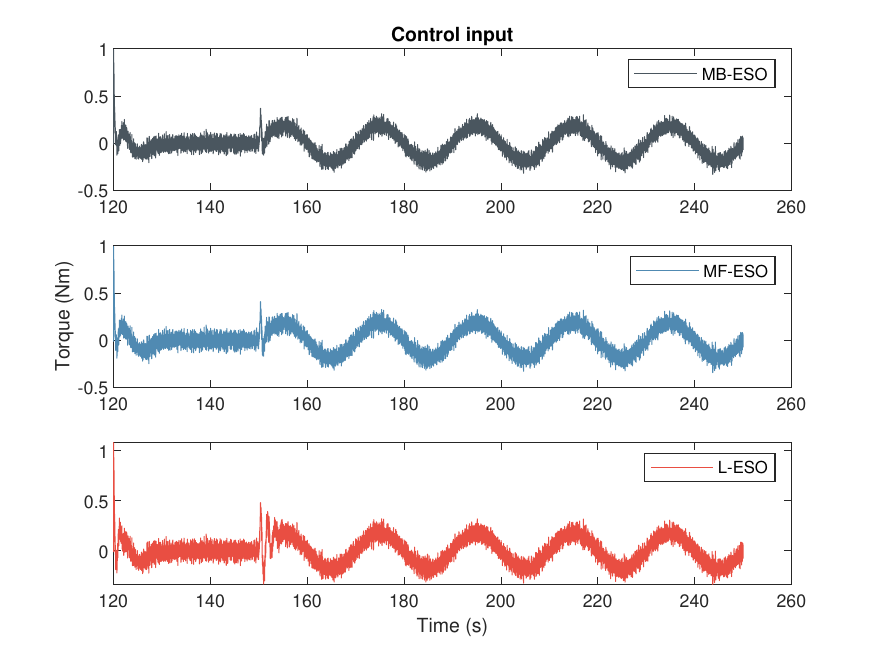}
    \caption{Control signal for MB-ESO, MF-ESO, L-ESO plotting from 120s.}
    \label{fig:control}
\end{figure}

The tracking performance and the control input are shown in Fig. \ref{fig:position} and Fig. \ref{fig:control}, respectively. %We have the following observations:
\begin{enumerate}
    \item MB-ESO and L-ESO have similar performance for the step reference tracking (see the zoom-in plot from 126 s to 134 s, Fig. \ref{fig:position}) after the training phase, see the position plot of $m_2$ in Fig. \ref{fig:position}, which are better than MF-ESO in terms of overshoot percentage (0 vs. $5$\textperthousand\ ) and settling time (12s vs. 16s).
    \item For external disturbance rejection (see the zoom-in plot from 170 s to 195 s, Fig. \ref{fig:position}), L-ESO's performance is the best. By re-visiting \eqref{eq:total-d-MF}, if the external disturbance has a linear component, a linear regression component can still capture it, e.g., the trends of going up and down in a sinusoidal external disturbance.
    
    %can learn the trend as the cost function also includes the external disturbance in Eq.\ref{eq-9}. Thus maintain the position of $m_2$ better than the other two methods.
    
    \item Adding external disturbance information to the observer can help reduce the required bandwidth. In our experiments, we found that MF-ESO and MB-ESO will need three times more bandwidth to achieve the same performance as the L-ESO.
    \item The control input of the L- ESO has more fluctuations compared with MF-ESO and MB-ESO, as shown in Fig. \ref{fig:control}. This is caused by the noise signal and the batch gradient descent method we choose to minimize the cost function. It can be smoothened by increasing the batch size in this example. %Overall, different learning approaches can be further investigated for improvement.

\end{enumerate}

\section{Hardware Experiments Results}
\label{sec:hardware}

\begin{figure}[htbp]
    \centering \includegraphics[width=0.5\linewidth, angle=270]{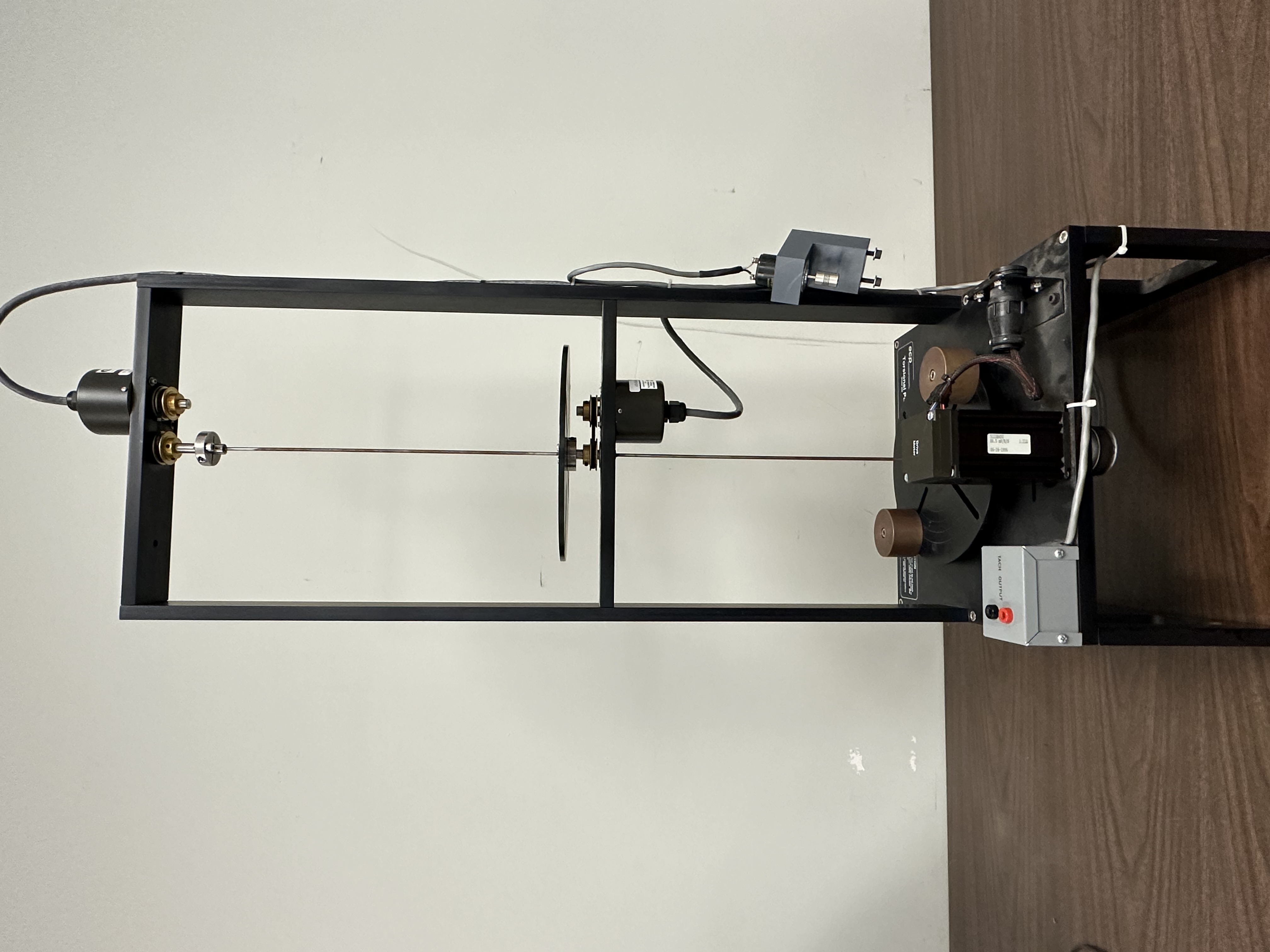}
    \caption{ECP Model 205 torsional testbed}
    \label{fig:plant}
\end{figure}

We conduct physical experiments on our ECP Model 205 torsional testbed \cite{ECP205}, see  Fig.\ref{fig:plant}. It is a mechanical system that consists of a flexible vertical shaft connecting two disks - a lower disk and an upper disk. Each disk is equipped with an encoder for position measurement. A DC servo motor drives the lower disk through a belt and pulley system, which provides a 3:1 speed reduction ratio. The system can be used to study the vibration of a torsional two-mass-spring system.

% \begin{wrapfigure}{R}{0.2\textwidth}
% \centering
%     \includegraphics[width=0.2\textwidth,angle=270]{Images/plant.jpg}
%   \caption{ECP Model 205 torsional testbed}
%   \label{fig:plant}
% \end{wrapfigure}

A personal computer with MATLAB\textsuperscript \textregistered  Simulink Desktop Real-Time\texttrademark~installed is used for computation. The computer is also equipped with a four-channel quadrature encoder input card (NI-PCI6601) and a multi-function analog and digital I/O card (NI-PCI6221). These cards interface with the torsional plant Model 205 for real-time data acquisition and control. The quadrature encoder input card enables the computer to receive position and velocity data from the encoders on the disks of the plant. The multi-function analog and digital I/O card allows the computer to send control signals to the DC servo motor that drives the lower disk.

\subsection{System Model} 
Since the MB-ESO, as a baseline approach, needs the dynamics information, we first use MATLAB\textsuperscript \textregistered  System identification toolbox and get the transfer function: $G(s)=\frac{4.6 \times10^4}{s^4+1.901s^3+1683s^2+1812s+0.1032}$.

\subsection{ESO and Controller Design }
As this testbed is again a fourth-order dynamic system, the same ESO design pipeline shown before can be applied.
\subsection{Experiment Results}
%The hardware results of the proposed method with comparison to the model-free and model-based approach using MATLAB®are presented.
Tracking a desired trajectory for the upper disk is the control objective. A sinusoidal wave with a frequency of $\pi/2$ rad/s and an amplitude $0.5\pi$ is applied in the training phase of L-ESO. $\omega_c$ and $\omega_o$ are set to 90 rad/s and 40 rad/s, respectively. The control gain is $5.5\times10^4$. 
%Instead of a step reference, which is too aggressive and contains broad bandwidth that may excite the resonant mode of the system, 
A trapezoidal profile reference with the final value $\pi$ is used. %A video link to our hardware experiment recording is available\footnote[1]{https://youtu.be/h-4bwztfdlY}.%, which is wildly used in industry and less aggressive, is tested for these three methods for comparison.

\begin{figure}[htbp]
    \centering
    \includegraphics[width=0.8\linewidth]{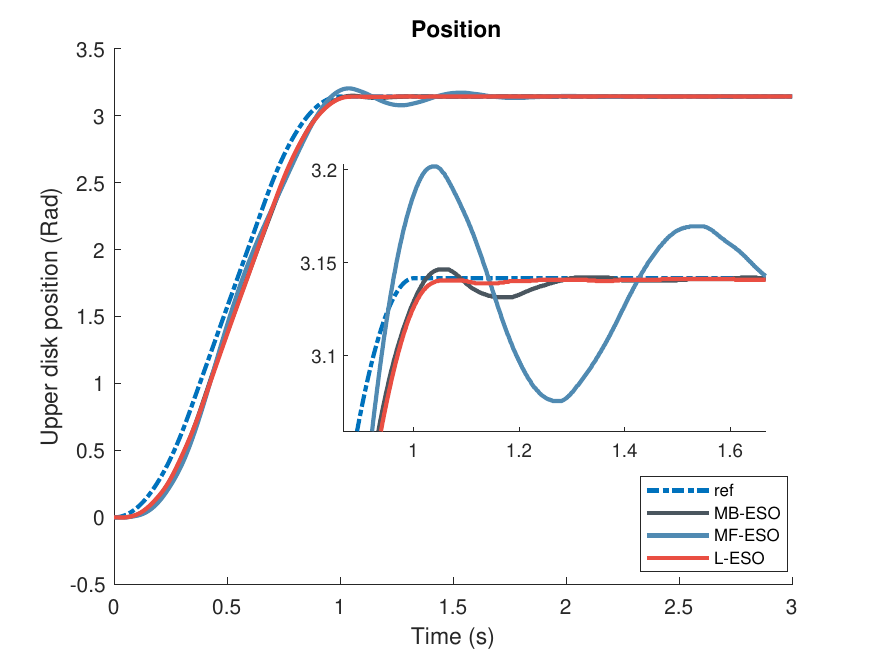}
    \caption{Upper disk position tracking: MB-ESO, MF-ESO, and L-ESO}
    \label{fig:hardwarepoistion}
\end{figure}

\begin{figure}[htbp]
    \centering
    \includegraphics[width=0.8\linewidth]{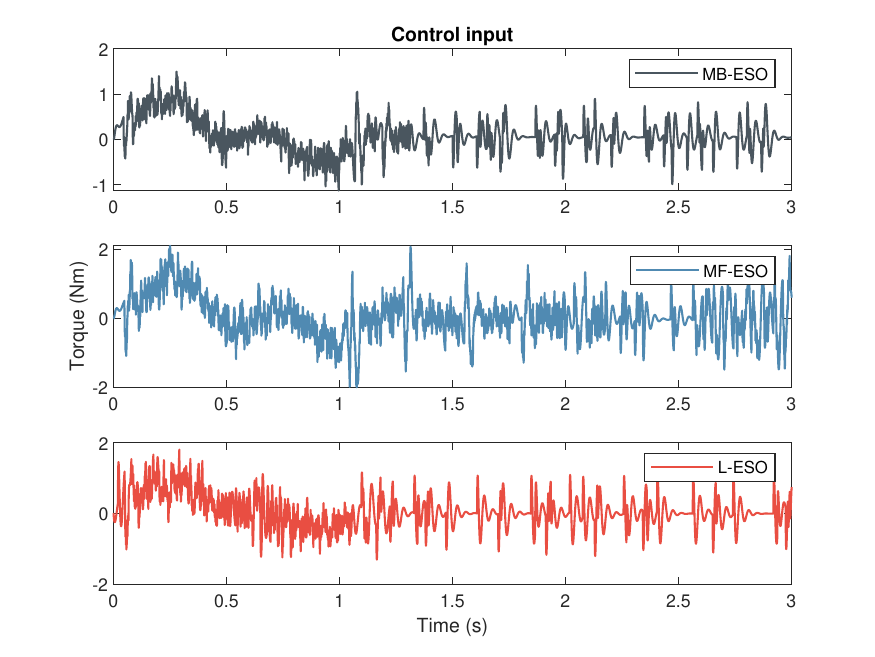}
    \caption{Control signal for MB-ESO, MF-ESO, L-ESO}
    \label{fig:hardwarecontrol}
\end{figure}
From the results illustrated in Fig. \ref{fig:hardwarepoistion} and Fig. \ref{fig:hardwarecontrol}, we have the following observations: 1) L-ESO has the best performance among all the methods after the training phase in terms of overshoot percentage and settling time. The reasons for L-ESO outperforming MB-ESO could be the imperfection of system identification or that our approach can learn internal as well as external disturbance. 2) The fluctuation of control input of L-ESO is between MF-ESO and MB-ESO, as shown in Fig. \ref{fig:hardwarecontrol}, which is different from the simulation result. This is because the learning rate is conservatively chosen due to the large noise in the hardware. Also, the trapezoidal profile reference is more smooth than the step reference, which is beneficial for learning.

\section{CONCLUSIONS}\label{sec:conclusion}
A novel learning-enabled extended state observer L-ESO with the capacity to memorize and generalize from past estimated disturbances is proposed in this paper. The machine learning model is seamlessly integrated into existing disturbance rejection control architecture as a flexible add-on for boosting robustness performance against unknown and time-varying disturbances. Compared with existing learning for control framework, our new paradigm does not rely on access to full states. In addition, the learning is guarded by disturbance rejection that provides an extra assurance layer to compensate for the imperfections of the machine learning model. The efficacy of the proposed approach has been supported by simulation and hardware experiments. In the future, we will further validate in real robotic testbeds.

\bibliographystyle{IEEEtran}
\bibliography{references}

% Generated by IEEEtran.bst, version: 1.12 (2007/01/11)
\begin{thebibliography}{10}
\providecommand{\url}[1]{#1}
\csname url@samestyle\endcsname
\providecommand{\newblock}{\relax}
\providecommand{\bibinfo}[2]{#2}
\providecommand{\BIBentrySTDinterwordspacing}{\spaceskip=0pt\relax}
\providecommand{\BIBentryALTinterwordstretchfactor}{4}
\providecommand{\BIBentryALTinterwordspacing}{\spaceskip=\fontdimen2\font plus
\BIBentryALTinterwordstretchfactor\fontdimen3\font minus
  \fontdimen4\font\relax}
\providecommand{\BIBforeignlanguage}[2]{{%
\expandafter\ifx\csname l@#1\endcsname\relax
\typeout{** WARNING: IEEEtran.bst: No hyphenation pattern has been}%
\typeout{** loaded for the language `#1'. Using the pattern for}%
\typeout{** the default language instead.}%
\else
\language=\csname l@#1\endcsname
\fi
#2}}
\providecommand{\BIBdecl}{\relax}
\BIBdecl

\bibitem{297904}
Y.~Hori, H.~Iseki, and K.~Sugiura, ``Basic consideration of vibration
  suppression and disturbance rejection control of multi-inertia system using
  {SFLAC} (state feedback and load acceleration control),'' \emph{IEEE
  Transactions on Industry Applications}, vol.~30, no.~4, pp. 889--896, 1994.

\bibitem{zhao2013active}
S.~Zhao and Z.~Gao, ``An active disturbance rejection based approach to
  vibration suppression in two-inertia systems,'' \emph{Asian Journal of
  control}, vol.~15, no.~2, pp. 350--362, 2013.

\bibitem{wang2024integrated}
Y.~Wang, L.~Dong, Z.~Chen, M.~Sun, and X.~Long, ``Integrated skyhook vibration
  reduction control with active disturbance rejection decoupling for automotive
  semi-active suspension systems,'' \emph{Nonlinear Dynamics}, pp. 1--16, 2024.

\bibitem{chen2021practical}
J.~Chen, Y.~Hu, and Z.~Gao, ``On practical solutions of series elastic actuator
  control in the context of active disturbance rejection,'' \emph{Advanced
  Control for Applications: Engineering and Industrial Systems}, vol.~3, no.~2,
  p. e69, 2021.

\bibitem{zheng2018active}
Q.~Zheng, Z.~Ping, S.~Soares, Y.~Hu, and Z.~Gao, ``An active disturbance
  rejection control approach to fan control in servers,'' in \emph{2018 IEEE
  Conference on Control Technology and Applications (CCTA)}.\hskip 1em plus
  0.5em minus 0.4em\relax IEEE, 2018, pp. 294--299.

\bibitem{Han2009}
J.~Han, ``From {PID} to active disturbance rejection control,'' \emph{IEEE
  Transactions on Industrial Electronics}, vol.~56, no.~3, pp. 900--906, 2009.

\bibitem{cui2017extended}
R.~Cui, L.~Chen, C.~Yang, and M.~Chen, ``Extended state observer-based integral
  sliding mode control for an underwater robot with unknown disturbances and
  uncertain nonlinearities,'' \emph{IEEE Transactions on Industrial
  Electronics}, vol.~64, no.~8, pp. 6785--6795, 2017.

\bibitem{zhang2021extended}
H.~Zhang, Y.~Li, Z.~Li, C.~Zhao, F.~Gao, F.~Xu, and P.~Wang,
  ``Extended-state-observer based model predictive control of a hybrid modular
  {DC} transformer,'' \emph{IEEE Transactions on Industrial Electronics},
  vol.~69, no.~2, pp. 1561--1572, 2021.

\bibitem{zhang2016active}
H.~Zhang, S.~Zhao, and Z.~Gao, ``An active disturbance rejection control
  solution for the two-mass-spring benchmark problem,'' in \emph{2016 American
  Control Conference (ACC)}.\hskip 1em plus 0.5em minus 0.4em\relax IEEE, 2016,
  pp. 1566--1571.

\bibitem{fu2016tuning}
C.~Fu and W.~Tan, ``Tuning of linear {ADRC} with known plant information,''
  \emph{ISA transactions}, vol.~65, pp. 384--393, 2016.

\bibitem{8689051}
Y.~Hui, R.~Chi, B.~Huang, and Z.~Hou, ``Extended state observer-based
  data-driven iterative learning control for permanent magnet linear motor with
  initial shifts and disturbances,'' \emph{IEEE Transactions on Systems, Man,
  and Cybernetics: Systems}, vol.~51, no.~3, pp. 1881--1891, 2021.

\bibitem{9736588}
J.~Wang, D.~Huang, S.~Fang, Y.~Wang, and W.~Xu, ``Model predictive control for
  {ARC} motors using extended state observer and iterative learning methods,''
  \emph{IEEE Transactions on Energy Conversion}, vol.~37, no.~3, pp.
  2217--2226, 2022.

\bibitem{zhang2022improving}
J.~Zhang and D.~Meng, ``Improving tracking accuracy for repetitive learning
  systems by high-order extended state observers,'' \emph{IEEE Transactions on
  Neural Networks and Learning Systems}, 2022.

\bibitem{kicki2022tuning}
P.~Kicki, K.~{\L}akomy, and K.~M.~B. Lee, ``Tuning of extended state observer
  with neural network-based control performance assessment,'' \emph{European
  Journal of Control}, vol.~64, p. 100609, 2022.

\bibitem{NeuralLander}
G.~Shi, X.~Shi, M.~O’Connell, R.~Yu, K.~Azizzadenesheli, A.~Anandkumar,
  Y.~Yue, and S.-J. Chung, ``Neural lander: Stable drone landing control using
  learned dynamics,'' in \emph{2019 International Conference on Robotics and
  Automation (ICRA)}, 2019, pp. 9784--9790.

\bibitem{guo2011convergence}
B.~Guo and Z.~Zhao, ``On the convergence of an extended state observer for
  nonlinear systems with uncertainty,'' \emph{Systems \& Control Letters},
  vol.~60, no.~6, pp. 420--430, 2011.

\bibitem{bai2019observers}
W.~Bai, S.~Chen, Y.~Huang, B.~Guo, and Z.~Wu, ``Observers and observability for
  uncertain nonlinear systems: A necessary and sufficient condition,''
  \emph{International Journal of Robust and Nonlinear Control}, vol.~29,
  no.~10, pp. 2960--2977, 2019.

\bibitem{chen2022relationship}
J.~Chen, Z.~Gao, Y.~Hu, and S.~Shao, ``A general model-based extended state
  observer with built-in zero dynamics,'' \emph{arXiv preprint
  arXiv:2208.12314}, 2023.

\bibitem{gao2003scaling}
Z.~Gao, ``Scaling and bandwidth-parameterization based controller tuning,'' in
  \emph{Proceedings of the 2003 American Control Conference, 2003.}\hskip 1em
  plus 0.5em minus 0.4em\relax IEEE, 2003, pp. 4989--4996.

\bibitem{wie1992benchmark}
B.~Wie and D.~S. Bernstein, ``Benchmark problems for robust control design,''
  \emph{Journal of Guidance, Control, and Dynamics}, vol.~15, no.~5, pp.
  1057--1059, 1992.

\bibitem{ECP205}
{Open AI}, ``Safety gym,''
  \url{http://www.ecpsystems.com/controls_torplant.htm} [Accessed: 3-23-2024].

\end{thebibliography}

%\addtolength{\textheight}{-12cm}   % This command serves to balance the column lengths
                                  % on the last page of the document manually. It shortens
                                  % the textheight of the last page by a suitable amount.
                                  % This command does not take effect until the next page
                                  % so it should come on the page before the last. Make
                                  % sure that you do not shorten the textheight too much.

%%%%%%%%%%%%%%%%%%%%%%%%%%%%%%%%%%%%%%%%%%%%%%%%%%%%%%%%%%%%%%%%%%%%%%%%%%%%%%%%

%%%%%%%%%%%%%%%%%%%%%%%%%%%%%%%%%%%%%%%%%%%%%%%%%%%%%%%%%%%%%%%%%%%%%%%%%%%%%%%%

\end{document}